\documentstyle[aps,multicol,epsf]{revtex}
\begin{document}
\draft

\title{Directed Surfaces in Disordered Media}

\author{A.-L. Barab\'asi$^*$, G. Grinstein, and M.A. Mu{\~n}oz}

\address{IBM Research Division, T.J.
Watson Research Center, P.O. Box 218, Yorktown Heights, NY 10598}
\date{\today}
\maketitle

\begin{abstract}
The critical exponents for a class of one-dimensional models of
interface depinning in disordered media can be calculated through
a
mapping onto directed percolation (DP).  In higher dimensions
these
models give rise to directed surfaces, which do not belong to the
directed percolation universality class.  We formulate a scaling
theory of directed surfaces, and calculate critical exponents
numerically, using a cellular automaton that locates the directed
surfaces without making reference to the dynamics of the
underlying
interface growth models.
\end{abstract}

\begin{multicols}{2}
\narrowtext

The last decade's intense theoretical, numerical, and
experimental
interest in the growth and roughening of interfaces has been
fueled in
part by the interdisciplinary aspects of the subject
\cite{rew}. Applications include fluid-fluid displacement
\cite{fluid}, imbibition in porous media \cite{imb},
and the motion of flux lines in superconductors \cite{sup}.  In
such
systems, an interface advances through a disordered (typically
porous)
medium, under the influence of an external driving force, $F$.
There
exists a critical value, $F_c$, of this force, such that for
$F<F_c$
the interface is pinned by the disorder, while for $F> F_c$ it
moves
with a constant velocity $v$.  In the vicinity of the depinning
transition ($F=F_c$), $v \sim f^\theta$, where $f = (F-F_c)/F_c$.
The
correlation length ($\xi$) characterizing the size of the pinned
regions in the plane of the interface diverges at $F_c$ as $\xi
\sim f^{- \nu}$.
For $F>F_c$ the width of the interface in steady state varies
with the
system size $L$
as $w(L) \sim L^\alpha$. Here $\theta$, $\nu$ and  $\alpha$ are
the
velocity, correlation length, and roughness exponents,
respectively.

In this paper we consider "directed percolation depinning" (DPD)
models \cite{imb,tl},
which are believed to describe the
depinning transitions in a variety of systems, among them
interfaces
described by the Kardar-Parisi-Zhang equation\cite{KPZ}  with
quenched noise
\cite{univ} and imbibition experiments\cite{imb}.
 The DPD model is  defined on a $d$-dimensional
lattice\cite{dim}, with periodic boundary conditions in the
($d-1$)
interface dimensions and open boundary conditions in the
dimension
corresponding to the direction of motion of the interface (the
$z$ direction, say).  Sites are randomly occupied by
impurities with probability $p$, and a fluid is imagined to push
its way upward from below.
At each time we randomly choose one of the impurity-free ``dry"
sites
neighboring the interface that separates the wet and dry regions.
The interface advances by ``wetting" the chosen site,
and {\it any site below it in the same column} (along the $z$
direction)
\cite{imb}.
In this model  only continuous, unbroken
surfaces of impurities that cover the entire (d-1)-dimensional
cross section of the lattice {\it
and contain no overhangs} can successfully block the interface
\cite{imb,tl}.
In 2d, this blocking surface is a collection of 1d lines, which
can be viewed as the backbone\cite{back}
of the infinite cluster of a
(1+1)-dimensional directed percolation (DP) problem that starts
from
one side of the lattice and percolates towards the other.
Thus the scaling exponents ($\alpha,
\theta, \nu$)
for the  DPD model in 2d can be obtained from the mapping
to the (1+1)-dimensional DP problem \cite{imb,tl}.  In
particular, the
correlation length of DPD can be identified with the longitudinal
correlation length of DP, giving $\nu = {\nu_\parallel}^{DP}
\approx 1.73$.  Similarly, one obtains the exponents $\alpha$ and
$\theta$ as
$\alpha={\nu_\perp}^{DP} / {\nu_\parallel}^{DP}
\approx 0.63$, and
$\theta={\nu_\parallel}^{DP} -{\nu_\perp}^{DP} \approx 0.63$,
where
${\nu_\perp}^{DP} \approx 1.10$
is the transverse correlation length exponent of DP.

While the DP theory correctly predicts all relevant exponents for
2d
DPD models, the mapping fails in higher dimensions. There is a
simple
topological reason for this: In (1+1)-dimensions, the directed
percolation backbone is a collection of 1d {\it lines}, capable
of
blocking the motion of the one-dimensional interface in 2d DPD.
In
3d, however, only an unbroken 2d {\it surface} of pinning sites
without overhangs can block the advance of the 2d interface.  The
collection of such surfaces forming the backbone of a blocking
cluster
in 2d (or in any other dimension) is referred to as a {\it
directed
surface} \cite{ds}.  Numerical measurements of the critical
exponents
characterizing the depinning transition of DPD models for $d>2$
\cite{amaral} confirm that directed surfaces belong to a
universality class different from DP.

In this paper we take several steps toward a description of the
DPD
depinning transition for $d>2$ in terms of
directed surfaces. First we introduce a deterministic two-state
cellular automaton (CA) that finds the directed surface for a
system
with an arbitrary distribution of impurities, and in
arbitrary
dimension.  The CA produces this surface  without making
reference to the dynamics of the underlying DPD interface growth
models.  Focusing on the
directed surfaces in this way allows us to formulate a scaling
theory
of the transition,
and thus describe this problem using the standard formalism of
critical phenomena.  Moreover, the CA allows us to measure
numerically
exponents not available from the DPD models.  Using  the
derived scaling laws we thereby
obtain a complete set of scaling exponents
characterizing the static properties of directed surfaces.

{\it Cellular automata and directed surfaces---} We first define
the
CA rules, and then explain why they generate the desired
directed surfaces. Consider a square 2d lattice in the $(x,z)$
plane.
At time $t=0$ each site $(x,z)$ is either independently occupied
by an
impurity ($s_0 (x,z)=1$) with probability $p$, or is empty ($s_0
(x,z) =0$).  The CA rule is defined as follows:
$s_{t+1}(x,z)=1$ if the following three conditions are
simultaneously
satisfied:

$s_t(x,z)=1$;

$s_t(x-1,z-1)+s_t(x-1,z)+s_t(x-1,z+1)>0$;

$s_t(x+1,z-1)+s_t(x+1,z)+s_t(x+1,z+1)>0$.

If any of these is not satisfied, then $s_{t+1}(x,z)=0$.
Boundary conditions are periodic in $x$ and open in $z$.
The CA rule
leaves untouched any occupied site that locally belongs to an
unbroken
1d path roughly perpendicular to the $z$ direction.  If an
occupied
site $(x,z)$ has an occupied neighbor or next near neighbor in
both
the $(x-1)$st and $(x+1)$st columns (see Fig. \ref{fig1}(a)),
then it is
part of such a path.  However, if the occupied site is at the
tip of a dangling branch (e.g., the site B in Fig \ref{fig1}(a)),
then at
the next time step it will be removed, since it is
missing a neighbor in one of the adjacent columns.
 Once B is removed, however, A is
left without a neighbor in one adjacent column, and so is itself
removed at the following time step.  In this fashion, all
dangling
branches and  isolated clusters of impurity sites  are
systematically eliminated.
Thus under successive applications of the rule, all impurities
that do {\it not}  belong to the backbone of the infinite cluster
of the
equivalent (1+1)-dimensional DP problem
(e.g., the shaded path in Fig. \ref{fig1}(a)), wherein $z$ and
$x$
respectively correspond to space and time, ultimately disappear.
If such an infinite DP cluster does not exist, then the fixed
point of the CA
has all sites empty.  Fig. \ref{fig1}(b)  shows the result of
applying this rule in a system of size $L=100$ with $p=0.55$.

The generalization to higher dimensions is straightforward.  We
discuss only the 3d case, defined on a cubic lattice with axes
labeled
$(x,y,z)$; the extension to $d>3$ is obvious.  In 3d,
$s_{t+1}(x,y,z)=1$ if the following five conditions are
simultaneously
satisfied:

$s_t(x,y,z)=1$;

$s_t(x-1,y,z-1)+s_t(x-1,y,z)+s_t(x-1,y,z+1)>0$;

$s_t(x+1,y,z-1)+s_t(x+1,y,z)+s_t(x+1,y,z+1)>0$;

$s_t(x,y-1,z-1)+s_t(x,y-1,z)+s_t(x,y-1,z+1)>0$;

$s_t(x,y+1,z-1)+s_t(x,y+1,z)+s_t(x,y+1,z+1)>0$.

If at least one of these conditions fails, then
$s_{t+1}(x,y,z)=0$.
Periodic and open boundary conditions apply in the
$(x,y)$ and $z$ directions, respectively.

As in 2d, for $d \geq 3$ the CA rule removes any site that does
not
belong to a locally continuous (d-1)-dimensional surface
perpendicular
to the $z$ direction.  It retains only unbroken surfaces, without
overhangs, capable of blocking the advance of an interface in the
associated DPD model.  It therefore locates {\it exactly all}
directed
surfaces in arbitrary dimension.  Thus one expects the critical
exponents measured for the DPD model and for the directed
surfaces
generated by the CA to coincide.  This conclusion is supported by
the
numerical results presented below.

{\it Scaling exponents for directed surfaces--} The final state
(fixed
point) of the CA depends on $p$. For small $p$ there are no
directed
surfaces in the system, so the average density of the final
state,
$\rho(p) = {1 \over L^d} \sum_{{\bf r}} s({\bf r})$, where ${\bf
r}={(i_1,i_2,.. i_d)}$, is zero.  Increasing $p$, one reaches a
critical value, $p_c$, so that for $p > p_c$, $\rho(p)$ is
nonzero. In
analogy with percolation, one expects that $\rho(p) \sim
(p-p_c)^{\beta}$.  The exponent $\beta$ is not known, and has not
previously been measured numerically, except in 2d, where the
directed
surface is the backbone of the infinite cluster of
(1+1)-dimensional
DP.  Since the exponent $\beta$ for this backbone is known to be
$2\beta^{DP}$ \cite{dhar}, where $\beta^{DP} \sim 0.28$ in (1+1)
dimensions, we have $\beta \approx 0.56$ in 2d.

To characterize further the directed surfaces for arbitrary $d$,
it is
helpful to define parallel (to the average orientation of the
surface) and perpendicular correlation lengths, $\xi_\parallel$
and
$\xi_\perp$, respectively.  Near the depinning transition they
diverge
as $\xi_\parallel \sim (p-p_c)^{-\nu_\parallel}$ and $\xi_\perp
\sim
(p-p_c)^{-\nu_\perp}$.

Finally, one can define exponents
associated
with the distribution of the sizes of voids or holes in the
directed
surface, i.e., empty regions totally surrounded by surface sites
\cite{amaral,SOD}.
The probability
distribution function, $P_v (v)$, for the void volumes $v$, and
$P_i
(s)$ for linear void sizes $s$ in the ith direction, are expected
to
behave algebraically: $P_v (v) \sim v^{-\tau_v}$, and, in 3d,
e.g.,
$P_{x ,y} (s) \sim s^{-\tau_\parallel}$, and $P_{z} (s) \sim
s^{-\tau_\perp}$.  As recent work by Huber {\it et al.} shows
\cite{greg}, in 2d
the void exponents $\tau_v$, $\tau_\parallel$, and
$\tau_\perp$ can be related to the exponents characterizing the
size
distribution of avalanches associated with the dynamics of
the DPD model
\cite{amaral} or its
self-organized version,
the ``self-organized depinning"
model \cite{SOD,mp}.

{\it Scaling theory--}  The first
advantage of describing
directed surfaces in static terms similar to those fruitfully
employed
for the
percolation problem
is that we can use the standard scaling arguments familiar
from critical phenomena to characterize the correlations and the
fractal nature of those surfaces\cite{aharony}.  As a first step,
we
derive a set of scaling relations for the critical exponents of
directed surfaces.  Imagine a coarse-grained density field $\psi
(\vec
x)$ for the surface.  Under rescaling of distances in the
directions
parallel and perpendicular to the surface via $x_{\parallel} = b
x'_{\parallel} $ and $x_{\perp} = b^{\alpha} x'_{\perp}$, the
field
$\psi$ and the distance $\Delta \equiv p-p_c$ from the critical
point
are assumed to rescale according to $\psi(x_{\parallel},
x_{\perp}) =
b^{\chi} \psi ' (x'_{\parallel}, x'_{\perp})$ and $\Delta =
b^{-1/{\zeta}} \Delta '$, respectively.  Here $b (>1) $ is the
length
rescaling factor, and $\alpha$, $\zeta$, and $\chi$ are critical
exponents.

The rescaling of lengths
 implies that the correlation length exponents
are given by
$\nu_{\parallel} = \zeta$, and $\nu_{\perp} = \alpha \zeta$:
Denoting the average density of the surface,
$<\psi(\vec x)>$, by $M$, it follows
 that $M$ vanishes like $\Delta^\beta$ as $\Delta \rightarrow
0^+$, with
$\beta = - \chi \nu_{\parallel}$.

To express $\chi$ in terms of the exponents for other correlation
functions, note that the scaling relations above imply
\begin{equation}
G(\vec x , \Delta) = \Delta^{-2\chi \nu_{\parallel}} h(
x_{\parallel}
\Delta^{\nu_{\parallel}} ,
x_{\perp}
\Delta^{\nu_{\perp}} )
\end{equation}
where
$G(\vec x  , \Delta) \equiv <\psi (\vec x) \psi (\vec 0 ) >$.
For $\Delta > 0$, $G$ has two parts, a disconnected piece
equal to $M^2$, and a connected piece $G_c (\vec x ,\Delta)
\equiv G(\vec x ,\Delta) - M^2$.
$G(\vec x ,\Delta)$ is proportional to the steady-state
probability
that both sites $\vec 0$ and $\vec x$ belong to the directed
surface,
i.e., to the product of
the probability
of site $\vec 0$ belonging to the directed
surface and the conditional probability
that $\vec x$ belongs to the directed surface, given that $\vec
0$ does.
This makes it clear that
$G (\vec x ,\Delta)$ vanishes like $M$ as $\Delta \rightarrow
0$, whereupon, close to $p_c$,
$G_c (x_{\parallel , \perp } ) \sim M x_{\parallel , \perp } ^
{-\beta / \nu_{\parallel , \perp}} $.  Using the standard
notation
of critical phenomena, we define the exponents
$\eta_{\parallel}$ and $\eta_{\perp}$ via $G_c (x_{\parallel ,
\perp} )
\sim M x_{\parallel , \perp} ^ {- ( d-2+\eta_{\parallel , \perp}
) }$,
leading to the scaling laws\cite{suscep}
\begin{equation}
\beta = \nu_{\parallel ,\perp} (d-2+\eta_{\parallel ,\perp}) ~~.
\label{scal1}
\end{equation}

The exponents $\eta_{\parallel}$ and $\eta_{\perp}$ are
straightforwardly related
to the fractal dimensions\cite{defD},
$D_{\parallel}$ and $D_{\perp}$, of the
directed surface
in the $\parallel$ and $\perp$ directions by\cite{hede}
$D_{\parallel} = 1-\eta_{\parallel}$ and $D_{\perp} = 3-d-
\eta_{\perp}$.  Moreover, the overall fractal
dimension, $D$, defined by the total
number of surface points within a distance
$R$ of a given point on the surface growing like $R^D$, is given
by
$D=2-\eta_{\perp}=d-\beta/\nu_{\perp}$.

Finally, for voids, we have the scaling relations
$\tau_\parallel =1+(\tau_v-1)(d-1+\alpha)$, and
$\tau_\perp=1+(\tau_v-1)(d-1+\alpha)/\alpha$.
In  2d the void exponents can be related to $\nu_{\perp}$,
$\nu_{\parallel}$, and $\beta$,
through the formula\cite{greg} $\tau_v -1 = (\nu_{\parallel} +
\nu_{\perp} -2\beta ) / (\nu_{\parallel} + \nu_{\perp} )$.

{\it Numerical results--} In 2d all the exponents defined here
are
available from the DP analogy.  For higher $d$,
$\nu_{\parallel}$,
$\nu_{\perp}$, and hence $\alpha$ have been obtained from DPD
simulations \cite{amaral}.  Others, such as $\beta$,
$\eta_{\parallel,
\perp}$ and $D_{\parallel , \perp}$ are difficult to compute from
DPD,
and so have not been measured.  The CA representation makes
numerical
determination of these rather straightforward.  Moreover, the
scaling
law (\ref{scal1}) shows that, given $\nu_{\parallel}$ and
$\nu_{\perp}$, one additional independent exponent suffices to
fix
all
the others, except the void exponents, which for $d>2$ have so
far
not
been related to the others.

In principle, the most straightforward exponent to measure
numerically
is $\beta$, obtained from plotting the density of the final state
of
the CA as a function of $p$.  In 2d the data show a fairly
unambiguous
scaling regime, yielding the value $\beta \approx 0.5$,
consistent
with the known value $\beta^{BB}=2\beta^{DP} \approx 0.56$.  In
3d,
near $p_c$, $\rho$ shows a rather precipitous jump that sharpens
with
increasing sample size, suggestive of either a first order phase
transition or a very small exponent $\beta$.  The data for
$\rho(p)$
are insufficient for one to choose between these possibilities.
We
therefore resort to determining $\beta$ by measuring
$D_\parallel$
and
$D_\perp$, and using the scaling relations
$D_\parallel=d-1-\beta/\nu_\parallel$ and
$D_\perp=1-\beta/\nu_\perp$,
and the values $\nu_\parallel=1.18 \pm 0.10$ and $\nu_\perp=0.57
\pm
0.05$, taken from DPD simulations\cite{amaral}.  The values for
$\beta$ thus obtained, together with our values of
$D_{\parallel}$
and
$D_{\perp}$, are listed in Table \ref{table}.  Our numerical
values
for
the void exponents are also given in the table.  The existence of
a fairly clear scaling regime for $D_\parallel$ with a value less
than 2 argues against the occurrence of a first-order phase
transition (as, of course, does the continuous phase transition
observed in the 3d DPD model).  The scaling observed in
the data of Fig. 2b  (spanning, for $P_v$,
more than two orders of magnitude in void sizes),
also
strongly supports the inference of a second-order transition.
 The 2d value for
$\tau_v$ is consistent with the prediction $\tau_v=1.80$ by Huber
{\it
et al.}  \cite{greg}.  Using the 3d numerical values, we estimate
from
the scaling relations above the value $\alpha \approx 0.48$, in
good
agreement with the value obtained from the DPD model.  This
provides
further confirmation that the directed surface generated by our
CA
indeed belongs in the same universality class as DPD.

One of the major benefits of the CA approach introduced in this
paper
is that it
replaces the dynamic models used to study the properties of
directed
surfaces with a static picture.  Dynamic models cannot capture
all directed surfaces existing in the system.  Models with random
updating, such as DPD, remove
small patches of the directed surface
in an uncontrolled fashion during large avalanches.
The
SOD models \cite{SOD} also systematically eliminate
branches of directed
surfaces. By contrast,
the CA is the first algorithm that identifies {\it all}
underlying directed surfaces,
which in turn determine
static exponents such as $\nu_\parallel,
\nu_\perp$, and $\alpha$, some of which (such as
$ \beta$), are not accessible from dynamic models.  The CA also
allows
direct calculation of the void-size exponents, from which, at
least in
2d, the avalanche
exponents of the dynamic models can be derived\cite{greg}.


We thank Sid Redner for helpful discussions.

\begin{figure}
\caption{(a) Schematic representation of the action of the CA
rule.
The final state of the CA contains only the backbone of the
underlying
DP cluster (solid circles).  Open circles represent dangling
branches or
isolated clusters that will eventually be eliminated.  (b)
Directed
surfaces produced by the CA rule in 2d with $p=0.55$ and
$L=100$.}
\label{fig1}
\end{figure}

\begin{figure}
\caption{Void size  distributions determined
numerically using the CA rule. (a) 2d: The upper, middle and
lower
curves correspond to $P_v(v)$, $P_x (s)$ and $P_z (s)$,
respectively.
We used $p=0.54$, $L=1000 \times 1000$, and averaged over 250
runs.
(b) 3d: The upper and lower curves correspond to $P_v(v)$, and
$P_z(s)$, respectively. In the middle we have two curves
superposed,
corresponding to $P_{x,y}(s)$. We used $p=0.74$, $L=100 \times
100
\times 100$, and averaged over 1500 runs.}
\label{fig3}
\end{figure}

\begin{table}
\caption{Exponents obtained from numerical simulations
using the CA rule, and from scaling relations.}
\begin{tabular}{||l|l|l||}
~~ Exponent ~~   & ~~2d  ~~~~~~~~~~~~~  &  ~~3d  ~~~~~~~~~~~~ \\
\hline
$\beta$    &    $0.5 \pm 0.07$        & $0.1 \pm 0.02$    \\
$D_{\parallel}$&$0.71\pm 0.03$       &  $1.9 \pm 0.05$    \\
$D_{\perp}$ &   $0.54 \pm 0.05$       & $0.8 \pm 0.1 $      \\
$\tau_v$   &    $1.7 \pm 0.1$        & $2.2 \pm 0.1$     \\
$\tau_x$   &    $2.2 \pm 0.1$        & $3.9 \pm 0.2$     \\
$\tau_z$   &    $2.8 \pm 0.2$        & $7.0 \pm 0.4$     \\
\end{tabular}
\label{table}
\end{table}

\end{multicols}

\begin{thebibliography}{99}

\bibitem[*]{byline} Present address: Department of Physics,
University of Notre Dame, Notre Dame, IN 46556-7062.

\bibitem{rew}
E.g., {\em Dynamics of Fractal Surfaces},
F. Family and T. Vicsek, eds. (World Scientific, Singapore,
1991);
P. Meakin, Phys. Rep. {\bf 235}, 189 (1993); T. Halpin-Healey and
Y.-C.
Zhang, Phys. Rep. {\bf 254}, 215 (1995); A.-L. Barab\'asi and
H.E.
Stanley,
{\em Fractal Concepts in Surface Growth} (Cambridge University
Press,
Cambridge, 1995).


\bibitem{fluid}
M.A. Rubio, C.A. Edwards, A. Dougherty, and J.P. Gollub, Phys.
Rev.
Lett. {\bf 63}, 1685 (1989); V.K. Horv\'ath, F. Family, and T.
Vicsek, J. Phys. A {\bf 24}, L25 (1991); Phys. Rev. Lett. {\bf
67},
3207 (1991); S. He, G.L.M.K.S.  Kahanda, and P.-z. Wong, Phys.
Rev.
Lett. {\bf 69}, 3731 (1992).

\bibitem{imb}
S.V. Buldyrev {\it et al.}, Phys. Rev. A {\bf 45}, R8313 (1992).


\bibitem{sup}
G. Blatter {\it et al.},
 Rev. Mod. Phys. {\bf 66},  1125 (1994).

\bibitem{tl}
L.-H. Tang and H. Leschhorn, Phys. Rev. A {\bf 45}, R8309 (1992).

\bibitem{KPZ} M. Kardar, G. Parisi, and Y.-C. Zhang,
 Phys. Rev. Lett. {\bf 56}, 889 (1986).



\bibitem{univ}
L.A.N. Amaral, A.-L. Barab\'asi, and H.E. Stanley,
Phys. Rev. Lett. {\bf 73}, 62 (1994); M. Kardar, L.-H. Tang,
and D. Dhar, Phys. Rev. Lett. {\bf 74}, 920 (1995).

\bibitem{dim} By a d-dimensional model we mean a
(d-1)-dimensional
interface moving in a d-dimensional system.


\bibitem{back} The backbone of an infinite  cluster
is defined as the part of the cluster from which all dangling
branches
or sub-clusters have been removed.


\bibitem{ds}
S.V. Buldyrev {\it et al.},
 Physica A {\bf 191}, 220 (1992).


\bibitem{amaral}
L.A.N. Amaral {\it et al.}, Phys. Rev. E {\bf 51}, 4655 (1995).

\bibitem{dhar}
B.M. Arora {\it et al.}, J.  Phys.  C {\bf 16}, 2913 (1983).

\bibitem{SOD}
S. Havlin {\it et al.}, in {\it Growth Patterns in Physical
Sciences and Biology} [Proc. 1991 NATO Advanced Research
Workshop,
Granada], ed. J.~M. Garcia-Ruiz {\it et al.},
(Plenum Press, New York, 1993); K. Sneppen, Phys. Rev. Lett. {\bf
69},
3539 (1992).

\bibitem{greg} G. Huber, M.H. Jensen, and K. Sneppen, Phys. Rev.
E
{\bf 52}, R2133 (1995).

\bibitem{mp}
S. Maslov and M. Paczuski, Phys. Rev. E {\bf 50}, R643 (1994);
Z. Olami, I. Procaccia, and R. Zeitak, Phys. Rev.  E {\bf 49},
1232
(1994).

\bibitem{aharony}  See, e.g., A. Aharony in {\it Directions in
Condensed Matter Physics}, G. Grinstein and G. Mazenko, eds.
(World Scientific, Singapore, 1986).

\bibitem{suscep}
More generally, $G_c(\vec x ,\Delta) \sim M x_{\parallel}^
{-(d-2+\eta_{\parallel})} f(x_{\perp} / x_{\parallel} ^ \alpha
)$,
as $\Delta \rightarrow 0$, where the scaling function $f(s)
\rightarrow 0$ as $s \rightarrow 0$, and $f(s) \rightarrow
s^{-(d-2+
\eta_{\parallel}) / \alpha} $
as $s \rightarrow \infty$.  It follows
that the ``susceptibility" $S$ defined by
$S (\Delta) \equiv \int d^d x G_c (\vec x ,\Delta)$
behaves as  $\Delta ^{\nu_{\parallel} (d-3+2\eta_{\parallel}
-\alpha) }$, whereupon the exponent $\gamma$ defined by
$S (\Delta) \sim \Delta^{-\gamma}$ is given by
$\gamma = \nu_{\parallel} (3+\alpha-2\eta_{\parallel}-d)$.

\bibitem{defD}
At $p=p_c$ the average number of points
on the surface contained in a $(d-1)$-dimensional hypersphere
of radius $R$, centered at an arbitrary point on the surface
and lying in the $(d-1)$-dimensional
parallel subspace, is expected to grow
like $R^{D_{\parallel}}$ for large $R$.
An analogous definition holds for $D_{\perp}$.

\bibitem{hede}
B. Hede, J.  Kert\'esz, and T. Vicsek, J. Stat. Phys. {\bf 64},
829 (1991).

\end{thebibliography}
\end{document}